\documentstyle[preprint,aps]{revtex}
\begin{document}
\draft
\preprint{SISSA/ISAS/53/EP}
\title{Intrinsic quasiparticle's spin and fractional quantum Hall
effect on Riemann surfaces}
\author{Dingping Li}
\address{International
School for Advanced Studies, SISSA,
I-34014 Trieste, Italy }
\date{\today}
\maketitle
\begin{abstract}
We derive the braid relations of the charged anyons
interacting with a magnetic field on  Riemann surfaces.
The braid relations are used to
calculate the quasiparticle's spin in the fractional quantum Hall
states on Riemann surfaces.
The quasiparticle's spin is found to be topological independent
and satisfies physical restrictions.
\end{abstract}

\narrowtext

The possibility of fractional statistics on two dimensional
surfaces was discovered in  Ref.\cite{lei}.
When two fractional-statistics particles (anyons)
are interchanged, the wave function changes by a phase
$\exp(i\theta)$, where $\theta$ is neither $\theta =0$ (Bose
statistics) nor $\theta =\pi$ (Fermion statistics).
The quasiparticles in fractional quantum Hall systems
(for a review on the
fractional quantum Hall effect (FQHE), see
Ref.\cite{girvin}) are
anyons \cite{arovas,halp} and this picture had been used to
construct the hierarchical wave function in the
FQHE \cite{halp}.
Anyons may also find applications in some condensed
matter systems \cite{sup}.
On higher dimension spaces ($D>2$), only Fermion and Boson
exist and the Fermion's spin is an half-integer,
the Boson's spin is an integer.
It will be very interesting to know
what is the spin of anyons
and the spin-statistics relation of anyons.
The spin-statistics relation of anyons is generalized one
if the spin $s$ equals to $s={\case\theta/2\pi}$.
Anyons in various models, for example, in
non-linear sigma models, Chern-Simons field theories and
relativistic quantum field theories on $2D$ dimension
spaces indeed have the generalized spin-statistics
relation \cite{rdt,fro}.
Naturally, we will ask the question
what is the spin-statistics relation of the quasiparticle
in the FQHE.
The recent discussion about the
quasiparticle's spin (QPS)
can be found in Refs.\cite{lidp,wen,kiv}.
Reference.\cite{lidp}
calculated the QPS by analyzing
the hierarchical wave function or by calculating
the Berry phase of the quasiparticles moving in a closed
path on the sphere.
Reference.\cite{wen} obtained the QPS
by analyzing the Ginzburg-Laudau-Chern-Simons (GLCS)
theory of the FQHE on the sphere.
On the other hand,
Reference.\cite{kiv} calculated the QPS
based on the GLCS theory on the disc geometry. However,
the results in Refs.\cite{lidp,wen,kiv} are
different from each other.  The ambiguity of
the QPS in the literature is
due to lack of a good definition of the QPS.

In this paper, we will calculate the QPS
by using braid relations of
anyons on Riemann surfaces
\cite{wu,thou,ein,wend,eina,imbo,bla}.
We will show that the QPS calculated in the following
is consistent with
physical restrictions and is intrinsic in the sense of
its topological independence.
The results of the present paper is agreed with the results
of Ref.\cite{lidp}.

Let us consider $N$ spinning anyons on an oriented compact
Riemann surface
with $g$ handles\cite{eina,imbo,bla}.
To define the anyon's spin,
we attach an oriented local frame to every particle.
When a particle moves on a curved surface (the torus is
a flat surface),
the attached frame is parallel transported and a
path-dependent frame-rotation is associated with the particle
transport.  Let us denote
the clockwise $2\pi$ rotation of the frame attached to the
particle by $R_{2\pi}$. The action of the operator
$R_{2\pi}$ on the wave function will give a phase
$\exp (i2\pi S)$. We define $S$ as the spin of the particle.
$S$ is equal to $1/2$ for the electron by this definition.
The braiding operators are
$\sigma_i ,\, \rho_{n,i} ,\, \tau_{n,i}$,
where $\sigma_i$  interchanges (clockwise)
particle $i$ and   particle $i+1$ and
$\rho_{n,i} ,\, \tau_{n,i}$ take
particle  $i$ around noncontractable
loops on the $m^{,}th$ handle.
Here we use the same definition of operators
$\rho_{n,i} ,\, \tau_{n,i}$
as Ref. \cite{eina}. The braid relations
for spinning anyons
on the  Riemann surface  with $g$ handles are
\begin{mathletters}
\label{brarel}
\begin{equation}
\sigma_j\sigma_{j+1}\sigma_j
=\sigma_{j+1}\sigma_j\sigma_{j+1} ,
\label{brarel:1}
\end{equation}
\begin{equation}
\tau_{m,j+1}= \sigma^{-1}_j \tau_{m,j}\sigma_j ,
\rho_{m,j+1}=\sigma^{-1}_j\rho_{m,j}\sigma_j ,
\label{brarel:2}
\end{equation}
\begin{equation}
\rho^{-1}_{m,j}\tau_{m,j+1}\sigma^{-2}_j\rho_{m,j}
\sigma^2_j\tau^{-1}_{m,j+1}=\sigma^2_j ,
\label{brarel:3}
\end{equation}
\end{mathletters}
\begin{equation}
\sigma_1 \sigma_2 \cdots \sigma^2_{N-1}\cdots \sigma_2
\sigma_1=R^{2(g-1)}_{2\pi}
\prod^g_n\rho^{-1}_{n,1}\tau^{-1}_{n,1}\rho_{n,1}\tau_{n,1}.
\label{relation}
\end{equation}
For the spinless anyons on the sphere, Equation\ (\ref{relation})
becomes $\sigma_1 \sigma_2 \cdots \sigma^2_{N-1}\cdots \sigma_2
\sigma_1=1 $, which had been derived in Ref. \cite{thou}.
It expresses the fact that a closed (clockwise) loop of particle
$1$ around all the other particles can be continuously shrunk to a
point on the rear side of the sphere.  Eq.\ (\ref{relation})
is the generalization of the case of the spinless anyons on the sphere
to the spinning anyons on
general Riemann surfaces\cite{eina,imbo}.
When we deform the loop of the left side of Eq.\ (\ref{relation})
to the loop of the right side of Eq.\ (\ref{relation}), described by
$\prod^g_n\rho^{-1}_{n,1}\tau^{-1}_{n,1}\rho_{n,1}\tau_{n,1}$,
the attached spin frame of  is rotated
($4\pi (g-1)$ rotation),
we obtain a phase $R^{2(g-1)}_{2\pi}$.  We need to include
this phase in the right side of the equation.
However for the charged
anyons in magnetic field, which is the case of
quasiparticles in FQHE,
the braid relation  \ (\ref{relation}) should be changed.
We also need to include
the Aharonov-Bohm phase $\exp (2\pi iq\Phi )$
in the right of
Eq.\ (\ref{relation}) because the charged anyon interacts with the
magnetic field, where
$q$ is the anyon's charge   and
$\Phi$ is the magnetic flux out of the surface.
Thus in stead of Eq.\ (\ref{relation}), for the charged anyons
on the magnetic field, we have
\begin{eqnarray}
\sigma_1 \sigma_2 \cdots \sigma^2_{N-1}\cdots \sigma_2
\sigma_1 & = &
\exp (2\pi iq\Phi ) R^{2(g-1)}_{2\pi} \nonumber \\
& \times & \prod^g_n\rho^{-1}_{n,1}
\tau^{-1}_{n,1}\rho_{n,1}\tau_{n,1} .
\label{relation1}
\end{eqnarray}
We will only consider the Abelian fractional statistics,
which means that
the representation of operator $\sigma_i$ is given by
$\sigma_i =\sigma =\exp{i\theta}{\bf 1}_M$,  where
${\bf 1}_M$ is  the $M \times M$ identity matrix.
Inserting
$\sigma_i =\sigma =\exp{i\theta}{\bf 1}_M$
in Eq.\ (\ref{brarel:2}) and
Eq.\ (\ref{brarel:3}), one obtains that
$\tau_{m,j}=\tau_m , \, \rho_{m,j}=\rho_m ,\,
\tau_m \rho_m =\sigma^2\rho_m \tau_m $.
These relations and Eq.\ (\ref{relation1}) yield
\begin{eqnarray}
\exp [2i(N-1)\theta] & = &
\exp [2\pi iq\Phi +4\pi(g-1)S]  \nonumber \\
& \times & \exp (-2ig\theta) .
\label{brarel3}
\end{eqnarray}
If there are several kinds of anyons, we need to introduce mutual
statistics\cite{mutual}.
The mutual statistics $\theta_{i,j}$ means that
when a particle of the $i^{,}{th}$ kind moves clockwise around
a particle of the $j^{,}{th}$ kind, we get a phase
$\exp (2\theta_{i,j})$.
$\theta_{i,i}=\theta_i$ actually is the
fractional statistics parameter of
the particle of the $i^{,}{th}$ kind.
The left hand of Eq.\ (\ref{brarel3})
is a phase which is obtained by moving one particle
(clockwise) around all other particles. If there exist
other kinds of particles,
instead of Eq.\ (\ref{brarel3}), we have
\begin{eqnarray}
\exp [2i(N_i-1)\theta_i
& + &
2i\sum_{j\not= i}^l N_j\theta_{i,j}]
=\exp (-2ig\theta_i) \nonumber \\
& \times &
\exp [2\pi iq_i\Phi +4\pi(g-1)S_i],
\label{mainre}
\end{eqnarray}
where
there are $l$ different kinds of  particles and
the spin  of the particle of the $i^{,}{th}$ kind
is $S_i$ and the charge is $q_i$.
Eq.\ (\ref{mainre}) gives a constraint
on the parameters (numbers, statistics and spin )
of anyons that
\begin{equation}
[((N_i-1+g)\theta_i
+  \sum_{j\not= i}^l N_j\theta_{i,j})
/ {\pi} ]
-q_i\Phi
 -  2(g-1)S_i
\label{const}
\end{equation}
is an integer.
Now we consider the FQHE on a  surface with $g$ handles.
We use the metric $ds^2=g_{z\bar z}dzd{\bar z}$
in complex coordinates.
The volume form is
$dv=[ig_{z\bar z} / 2]dz\wedge d{\bar z}
= g_{z\bar z}dx\wedge dy$.
As in the case of the FQHE on the disc, sphere and torus,
we apply a constant magnetic field on the surface.
The natural generalization
of the  constant magnetic field to high genus Riemann
surfaces\cite{bolte} is
$F=Bdv=(\partial_zA_{\bar z}-
\partial_{\bar z}A_z)dz\wedge d{\bar z}$.
Thus  $ \partial_zA_{\bar z}-
\partial_{\bar z}A_z =ig_{z\bar z}B / 2$.
The flux $\Phi$ is given by $2\pi \Phi = \int F =BV$,
where $V$ is the area of the surface and
we assume here $B>0$ ($\Phi >0$).
The Hamiltonian of an electron on the surface
under the magnetic field is given
by the Laplace-Beltrami operator,
\begin{eqnarray}
H & = & [1/ 2m \sqrt{g}]
(P_{\mu}-A_{\mu})g^{\mu \nu}\sqrt{g}(P_{\nu}-A_{\nu})
\nonumber \\
& = & [ g^{z\bar z} / m]
[(P_z-A_z)(P_{\bar z}-A_{\bar z}) \nonumber \\
& + &  (P_{\bar z}-A_{\bar z})(P_z-A_z)]  \\
\label{hamil}
& = & [2g^{z\bar z}/ m]
(P_z-A_z)(P_{\bar z}-A_{\bar z})+[B/ 2m] \nonumber
\end{eqnarray}
where $g^{z\bar z}=[1 / g_{z\bar z}]$ and
$P_{z}=-i\partial_z ,\, P_{\bar z}=-i\partial_{\bar z}$.
The inner product of two wave functions is defined as
$<\psi_1 | \psi_2 >=\int dv {\bar \psi_1 }\times \psi_2$.
$H^{\prime}=[2g^{z\bar z} / m]
(P_z-A_z)(P_{\bar z}-A_{\bar z})$
is a positive definite hermitian operator
Because $<\psi | H^{\prime} |\psi > \, \geq 0$ for any $\psi$.
Thus if  $H^{\prime} \psi =0$,
$\psi$ satisfies
$(P_{\bar z}-A_{\bar z})\psi =0$.
The solutions of this equation
are the ground states of the Hamiltonian $H$ or $H^{\prime}$,
or the lowest Landau levels (LLL).
The existence of the solutions of this equation
is guaranteed by Riemann-Roch theorem\cite{griff}.
The solutions belong to the
holomorphic line bundle under the gauge field.
Riemann-Roch theorem tells us that
$h^0(L)-h^1(L)=deg(L)-g+1$,
where $h^0(L)$ is the dimension of the holomorphic line bundle
or the degeneracy of the ground states of the Hamiltonian $H$,
$h^1(L)$ is the dimension of the holomorphic differential and
$deg(L)$ is the degree of the line bundle which is equal to
the first Chern number of the gauge field, or the magnetic flux
out of the surface, $\Phi$.
Because $deg(L) >2g-2$ (the magnetic field is very strong
in the FQHE), $h^1(L)$ is equal to zero \cite{griff}
and $h^0(L)=\Phi-g+1$.
As a consistent check,  $h^0(L)$ indeed
gives the right degeneracy
of the ground states in the case of
a particle on the sphere and torus interacting
with a magnetic-monopole field.
In the case of high genus surfaces,
Ref.\ \cite{avron} had discussed
the degeneracy of the LLL for the
leaky tori (see also Ref.\ \cite{asory} for the related
discussions).

If the filling factor is $1 /  m$ in the state of the FQHE
\cite{laughlin},
we suppose to have relation $m(N+\Delta_g)=\Phi$.
We assume that $\Delta_g$ is independent of $m$, and
this assumption will be justified by
the explicit constructions
of some examples of Laughlin wave functions on high
genus surfaces \cite{iengoli}.
If $m=1$, we have an integer quantum Hall state and expect
that the LLL are completely filled. In this case,
$N=\Phi-\Delta_g$, and $N$  is also
the degeneracy of the LLL, which should be
equal to $\Phi-g+1$ (according to the discussions above).
Thus $\Delta_g$ is  equal to $g-1$.
The relation between the electron
numbers and the flux for the FQHE
at the filling factor as $1 /  m$ is then
$m(N+g-1)=\Phi$, which indeed gives correct results
in the case of
the Laughlin state on the sphere and torus \cite{haldane}.
If $N_q$ quasiparticles are created,
one has $m(N+\Delta_g)+N_q=\Phi$.
The mutual statistics between the electron and quasiparticle is
$2\theta_{mut}=2\pi$, the charge of the quasiparticle is
$1 / m$ and the statistics parameters is $\pi / m$.
We remark that, if $2\theta_{i,j}=2\pi, i\not= j$ in
Eq.\ (\ref{const}) or Eq.\ (\ref{mainre}), we can simply
omit the term $N_j\theta_{i,j}$.
By applying Eq.\ (\ref{const}) to  those $N_q$ quasiparticles,
one can show that
\begin{equation}
[(N_q-1+g)/
m]+N- [\Phi / m] -2(g-1)S_q
\label{spin1}
\end{equation}
is an  integer.
{}From Eq.\ (\ref{spin1}) and
the equation $m(N+\Delta_g)+N_q=\Phi$,
the QPS turns out to be
$S_q=[1 / 2m]+[n / 2(g-1)]$
($n$ is an integer and $g \not= 1$ is assumed).
Let us discuss how to fix $n$.
We consider a cluster of particles which contains
$n_i$ particles of the $i^{,}{th}$ kind with
the mutual statistics
$\theta_{i,j}$. By using the method developed
in Ref.\ \cite{thou},
we get the statistics,  spin
and charge of the cluster,
\begin{equation}
S_c  = \sum_i [n_i(n_i-1)\theta_i / 2\pi ]
+n_iS_i +
\sum_{i \not= j} n_in_j\theta_{i,j} / 2\pi
\label{culster:2}
\end{equation}
and $q_c=\sum n_iq_i$ and $\theta_c=\sum_i n_i^2\theta_i+
\sum_{i \not= j}n_in_j\theta_{i,j}$
are the charge and statistics of the cluster respectively.
If the cluster's charge is an odd (even) integer
and the cluster satisfies the
fermionic (bosonic) statistics, we suppose that the cluster
contains only an  odd (even) number of electrons
(for example, see Ref.\ \cite{frozee}),
and thus the cluster's spin
is an half-integer (integer). If the cluster contains
$m$ quasiparticles in the above example, the  cluster's charge
is $1$ and its statistics of is fermionic. This
cluster shall be the hole of the electron
and the cluster's spin
is an half-integer (see also Ref.\ \cite{fro}).
By using Eq.\ (\ref{culster:2}) for this cluster, we get
a restriction for the QPS,
$[mn / 2(g-1)]=integer$. If $m$ and $g-1$ are coprime to each
other (for example, $g$ is equal to $0,2,3$),
$n$ must be equal to $2n^{\prime}(g-1)$
($n^{\prime}$ is an integer).
Thus $S_q$ is equal to $1 / 2m$ (up to an integer) and the
spin-statistics relation is the standard one.
However, when $m$ and $g-1$ are not coprime to each other
for some high genus surfaces, there exist other solutions
for the spin except $1 / 2m$.
We write $m=kp$ and $g-1=kq$, where
$p$ and $q$ are coprime to each other,
we have solutions $n=2n^{\prime}q$.
The other solutions of the  QSP is
$S_q=[1 / 2m]+[n^{\prime}/  k]$
where $n^{\prime}=1,\cdots , k-1$.
Therefor $S_q$  can not be completely
fixed by using the braid group analysis.
However, we shall point out that
it is highly unlikely that
those {\bf other} solutions are
the {\bf true} QPS,  as
we expect that the value of the spin shall be intrinsic
and does not depend on the surface where quasiparticles live.
To completely fix the QPS  on Riemann surfaces,
we can obtain the QPS by analyzing the wave function of the
quasiparticles on Riemann surfaces
(we plan to do so elsewhere),
as it was done  for the case on the sphere\cite{lidp}.

Let us calculate the QPS in
the standard hierarchical
state\cite{halp,lidp,haldane,blok,read,gread,lidpt}.
We remark that the above method can be used to
calculate the QPS in other kinds of quantum Hall fluids,
for example, the multi-layered FQHE or Jain state\cite{jain} etc..
The hierarchical state is described by a symmetric matrix
$\Lambda_{i,j}, i,j=1,2,\cdots , l$, where
$\Lambda_{i,i+1}=\Lambda_{i+1,i}=\pm 1$,
$\Lambda_{1,1}$ is an odd integer  and $\Lambda_{i,i}, i\not= 1$
are even integers, where  $l$ is the level of the
hierarchical state and
$N_i$ is the number of the
particles in level $i$ ($N_1$ is the number of the electrons,
$N_2$ is the number of the condensed quasiparticles (or holes)
of the first level (Laughlin) state, etc.).
On the torus\cite{haldane,lidpt},  we have a relation,
$\sum_j \Lambda_{i,j}N_j=\delta_{i,1}\Phi$, and on the
sphere\cite{lidp,haldane}, the relation is
$\sum_j \Lambda_{i,j}N_j -\Lambda_{i,i}=\delta_{i,1}\Phi$.
Following the discussion about the Laughlin state
on Riemann surfaces, we expect that the relation is
$\sum_j \Lambda_{i,j}N_j + (g-1) \Lambda_{i,i}=
\delta_{i,1}\Phi$ for the hierarchical state
on Riemann surfaces.
We define a $l$ dimension integer
lattice with bases $E_i$ and the inner products
$E_i\cdot E_j =\Lambda_{i,j}$ (see Ref.\ \cite{lidpt})
The above equation
can be rewritten as
\begin{equation}
\sum_{i=1}^{l} N_iE_i + (g-1) (E_i\cdot E_i)E_i^{\star}=E_1^{\star}\Phi ,
\label{hierar:1}
\end{equation}
where $E_i^{\star}$ are the bases
of the inverse lattice $E_i$ and defined
by $E_i^{\star}\cdot E_j =\delta_{i,j}$.
It can be verified that
$E_i^{\star}\cdot E_j^{\star} =\Lambda_{i,j}^{-1}$.
The quasiparticle is described by a
vector ${\cal Q}_k =k_iE_i^{\star}$ ($k_i$ is an integer)
on the lattice $E_i^{\star}$ (see Refs.\ \cite{blok,read}).
The statistics parameter of this quasiparticle is
$\theta_k = {\cal Q}_k\cdot
{\cal Q}_k \pi=\sum_{i,j}k_i\Lambda_{i,j}^{-1}k_j$
and the charge is $Q_k={\cal Q}_k \cdot E_1^{\star}=\sum_ik_i
\Lambda_{i,1}^{-1}$. The mutual statistics between
the quasiparticles
${\cal Q}_{k}$ and ${\cal Q}_{k^{\prime}}$ is
$\theta_{k,k^{\prime}}=
{\cal Q}_{k}\cdot {\cal Q}_{k^{\prime}}\pi
=\sum_{i,j}k_i\Lambda_{i,j}^{-1}k^{\prime}_j$.
If $N$ quasiparticles
denoted by the vector ${\cal Q}_k$, is created,
one has  $\sum_{i=1}^{l} N_iE_i +
(g-1) (E_i\cdot E_i)E_i^{\star}+ N
{\cal Q}_k=E_1^{\star}\Phi$.
Making inner product on two sides of this equation with
${\cal Q}_k$,
one yields
\begin{equation}
{\frac {N\theta_k}{\pi}}+(g-1)\Delta \cdot {\cal Q}_k-
Q_k \Phi =integer,
\label{delt}
\end{equation}
where $\Delta = \sum_{i=1}^{l}(E_i \cdot E_i)E_i^{\star}$.
Applying Eq.\ (\ref{const}) to those quasiparticles and comparing
it with Eq.\ (\ref{delt}), one gets
\begin{equation}
S_k={\theta_k \over 2\pi}-{\frac {\Delta \cdot {\cal Q}_k} {2}}
+{\frac {n}{2(g-1)}}, \, n=integer.\, g\not= 1.
\label{result}
\end{equation}
By using the argument in Ref.\ \cite{thou}
(which we did for the quasiparticle
of the Laughlin state), we can fix $n$ in some cases.
The charge of the quasiparticle $E_i$ (denoted by the vector $E_i$)
is $\delta_{i,1}$ and the statistics is
$\theta = \delta_{i,1}\pi$.
Thus the spin of this quasiparticle is
${\frac {\delta_{i,1}} {2}}+integer$. Because
$E_i=\sum_j \Lambda_{i,j}E_j^{\star}$, the quasiparticle $E_i$
is a cluster which
contains  $\Lambda_{i,j}$ quasiparticles
with the vector as $E_j^{\star}$.
If $\det{\Lambda}$ and $g-1$ are coprime to each other,
by using Eq.\ (\ref{culster:2}) for the cluster,
we find that
$S_i={\frac {E_i \cdot E_i} {2}}-
{\frac {\Delta \cdot E_i} {2}}-{1\over 2}$
for the quasiparticle $E_j^{\star}$ when $l-i$
is an even integer and
$S_i={\frac {E_i \cdot E_i} {2}}-
{\frac {\Delta \cdot E_i} {2}}$ when $l-i$ is an odd integer.
We will use notation $i \in i_{e}$
if $l-i$ is an even integer.
Generally, for the  quasiparticle
${\cal Q}$, we find that
\begin{equation}
S_{\cal Q}={\frac { {\cal Q}\cdot {\cal Q}} {2}}-
{\frac {\Delta \cdot {\cal Q}} {2}}-{1\over 2}\Delta^{\prime} \cdot
{\cal Q},
\label{result:1}
\end{equation}
where  $\Delta^{\prime}=\sum_{i} E_i, i\in i_e$.
If the quasiparticles have the standard spin-statistics relation,
it is required that ${\frac {\Delta \cdot {\cal Q}} {2}}
+{1\over 2}\Delta^{\prime} \cdot {\cal Q}$ is an integer.
Indeed, this number is always an integer for the Laughlin state.
However, in the hierarchical state, this number may not be an integer.
Thus the quasiparticles in the hierarchical state usually do not
have the standard spin-statistics relation.

If $\det{\Lambda}$ and $g-1$ are not coprime to each other,
there exist other solutions, not only Eq.\ (\ref{result:1}).
As we argued in the case of the Laughlin state, these
{\bf other} solutions is {\bf unlikely}
the {\bf true} QPS. Thus
we suppose that Eq.\ (\ref{result:1}) is
the  spin for quasiparticles and it is topological independent.
The above method  does not give any
information about the QPS
in the FQHE on the torus ($g=1$).
However, the above discussion
strongly suggests that the QPS in the FQHE on the torus
is also given by Eq.\ (\ref{result:1}) which suppose
to be the QPS on any Riemann surfaces.

The Lagrangian for the long-distance physics of the Hall fluid
on Riemann surfaces is \cite{wen},
\begin{eqnarray}
{\cal L}& = & {1\over 4\pi}(\alpha_{\mu , i}\Lambda_{i,j}\epsilon^{\mu
\nu \lambda}\partial_{\nu}\alpha_{\lambda , j}
 + 2A_{\mu}t_i\epsilon^{\mu \nu \lambda}
\partial_{\nu}\alpha_{\lambda , i} \nonumber \\
& + & 2\omega s_i\epsilon^{0 \nu \lambda}
\partial_{\nu}\alpha_{\lambda , i}),
\end{eqnarray}
where $\omega$ is the connection one form (the curvature
is given by $R=d\omega$).  In the case of the hierarchical state,
$t_i$ is equal  to $\delta_{i,1}$
and the matrix $\Lambda$ is one  we gave in the previous discussion.
By using Eq.\ (\ref{hierar:1}), we can show that
$s_i=\Lambda_{i,i}$. So $s_i$ is a topological
independent constant. It is reasonable to believe that
$s_i$ is  a topological independent constant for any kinds
of quantum Hall fluids.
Due to the presence of the third term in the
Lagrangian,
the spin-statistics relation usually is not
generalized one \cite{wen}.

The QPS in some fractional quantum Hall
states on Riemann surfaces has been  calculated
by using the braid relation for spinning anyons.
The value of the  QPS obtained satisfies
physical restrictions and
is found to be topological independent.
Riemann-Roch theorem has been used in the discussions
about the FQHE on Riemann surfaces.
Because the methods used in this paper are
rather general,  they  can be used to calculate
the particles's spin in other quantum Hall fluids,
chrial spin fluids and anyon superfluids.

\acknowledgments

I would like to thank Professor S. Cecotti,
Professor B. Dubrovin, T. Einarsson,
and  especially Professor R. Iengo
for enlightening discussions.
I also thank Professor J.M. Leinaas for discussions and
hospitality at the Physics Department  of Oslo University
(Norway).

\end{document}